\documentclass[aps,prl,twocolumn,superscriptaddress,a4paper,english,longbibliography]{revtex4-2}
\usepackage{times}
\usepackage{color}
\usepackage[colorlinks=true,urlcolor=blue,citecolor=blue]{hyperref}
\usepackage{url}
\usepackage{breakurl}
\usepackage{soul}
\usepackage{graphicx}
\usepackage{amsmath}
\usepackage{subfigure}
\usepackage{xcolor}
\usepackage{amssymb}
\usepackage{latexsym}
\usepackage{hyperref}
\DeclareGraphicsExtensions{.png,.eps}
\usepackage{float}
\usepackage{appendix}
\usepackage{etoolbox}

\usepackage{xcolor}
\usepackage[urlcolor=blue]{hyperref}      
\hypersetup{
	colorlinks = true,                    
	citecolor = {blue},
	linkcolor = {purple},
}

\begin{document}	
	
\title{Automatically Differentiable Quantum Circuit for Many-qubit State Preparation}

\author{Peng-Fei Zhou}
\affiliation{Department of Physics, Capital Normal University, Beijing 100048, China}
\author{Rui Hong}
\affiliation{Department of Physics, Capital Normal University, Beijing 100048, China}
\author{Shi-Ju Ran} \email[Corresponding author. Email: ] {sjran@cnu.edu.cn}
\affiliation{Department of Physics, Capital Normal University, Beijing 100048, China}
\date{\today}

\begin{abstract}
	Constructing quantum circuits for efficient state preparation belongs to the central topics in the field of quantum information and computation. As the number of qubits grows fast, methods to derive large-scale quantum circuits are strongly desired. In this work, we propose the automatically differentiable quantum circuit (ADQC) approach to efficiently prepare arbitrary quantum many-qubit states. A key ingredient is to introduce the latent gates whose decompositions give the unitary gates that form the quantum circuit. The circuit is optimized by updating the latent gates using back propagation to minimize the distance between the evolved and target states. Taking the ground states of quantum lattice models and random matrix product states as examples, with the number of qubits where processing the full coefficients is unlikely, ADQC obtains high fidelities with small numbers of layers $N_L \sim O(1)$. Superior accuracy is reached compared with the existing state-preparation approach based on the matrix product disentangler. The parameter complexity of MPS can be significantly reduced by ADQC with the compression ratio $r \sim  O(10^{-3})$. Our work sheds light on the ``intelligent construction'' of quantum circuits for many-qubit systems by combining with the machine learning methods.
\end{abstract}

\maketitle


\textit{Introduction.---} Quantum states with entangled qubits play a fundamental role in the field of quantum information and computation~\cite{AJPNMAQCQI2002}. In many cases such as quantum encrypted communications~\cite{PhysRevLett.69.2881,PRLMKWHCodingQC1996} and measurement-based quantum computations~\cite{NatureMeasurementbased2009}, a key step is to prepare the desired states on the quantum platform. Such tasks can be done by deriving the quantum circuits that transforms the initial states, which are usually product states with no entanglement, to the target states (see, for instance, \cite{PRLBCHRsatePreparation2001,PRLRKJPreparationstate2002,PRAQPMBPreparationstate2011,PhysRevA.101.032310,araujo_divide-and-conquer_2021}).

For the existing quantum platforms such as superconducting circuits~\cite{RevModPhys.73.357}, cold atoms~\cite{PhysRevLett.97.200405}, and photonic interferometries~\cite{PhysRevA.52.3489,PhysRevLett.75.4710}, certain restrictions should be imposed to the quantum circuit to make it realizable. For instance, a quantum circuit usually represents a unitary transformation formed by the unitary gates acted on a small number of qubits. Methods have been proposed to compile an arbitrary two-qubit gate to the combination of certain elementary gates, such as CNOT gates and single-qubit rotations~\cite{PRABarennoEGQC1995,PRADivtwobitqc1995}. It is also a hot topic to develop non-Hermitian quantum computation schemes by quantum walk~\cite{PhysRevLett.102.065703}.

Quantum circuits can be derived by, e.g., implementing Schmidt decompositions on the target state~\cite{PRAQPMBPreparationstate2011}. In recent years, the number of qubits in the quantum platforms increases in an extremely fast speed~\cite{HM17QS2017,AABB+2019GoogleQ}. This raises more challenges on constructing the circuits, since the dimension of the Hilbert space, where the states and operators are defined, grows exponentially with the number of qubits. Developing efficient methods to derive large-scale quantum circuits becomes increasingly urgent. 

Tensor network (TN) provides a powerful mathematical tool in simulating the quantum models with exponentially large Hilbert space~\cite{VMC08MPSPEPSRev2008,CV09TNSRev2009,O14TNSRev2014,RTPC+17TNrev2020,O19TNrev2019}. It reduces the complexity of representing the states therein to be polynomial to the system size under the constraint that the states satisfy the area laws of entanglement entropy~\cite{S93Sarea1993,ECP10AreaLawRev2010}. Recently, machine learning techniques have been introduced to the TN simulations and circuit constructions~\cite{PhysRevA.98.032309,PhysRevA.98.062324,Arrazola_2019,PhysRevLett.123.260404}. For instance, automatic differentiation, which has wide applications in machine learning for optimizing neural networks, is utilized to develop efficient TN algorithms, where the gradients of tensors can be obtained by back propagation~\cite{PRXDPTNwanglei2019,dTRG2020}. 

In this work, we propose the automatically differentiable quantum circuit (ADQC) for preparing the states that may contain large numbers of qubits. We introduce latent gates whose decompositions determine the unitary gates that form the circuit to satisfy the unitary constraints. The latent gates, which are automatically differentiable and are not imposed by any constraints, are updated by back propagation method to minimize the distance between the evolved and target states. ADQC is benchmarked on preparing the ground states of one-dimensional (1D) quantum spin chains and random matrix product states (MPS) as examples. For the Heisenberg and XY chains, ADQC surpasses the circuits constructed from the matrix product disentangler \cite{PhysRevA.101.032310}. The number of parameters can be significantly reduced representing the MPS with ADQC, where we have the compression rate $r = N_Lr_0$ with $N_L \sim O(1)$ the number of layers in the ADQC and $r_0 \sim O(10^{-3})$. 

\begin{figure}[tbp]
	\centering
	\includegraphics[angle=0,width=1\linewidth]{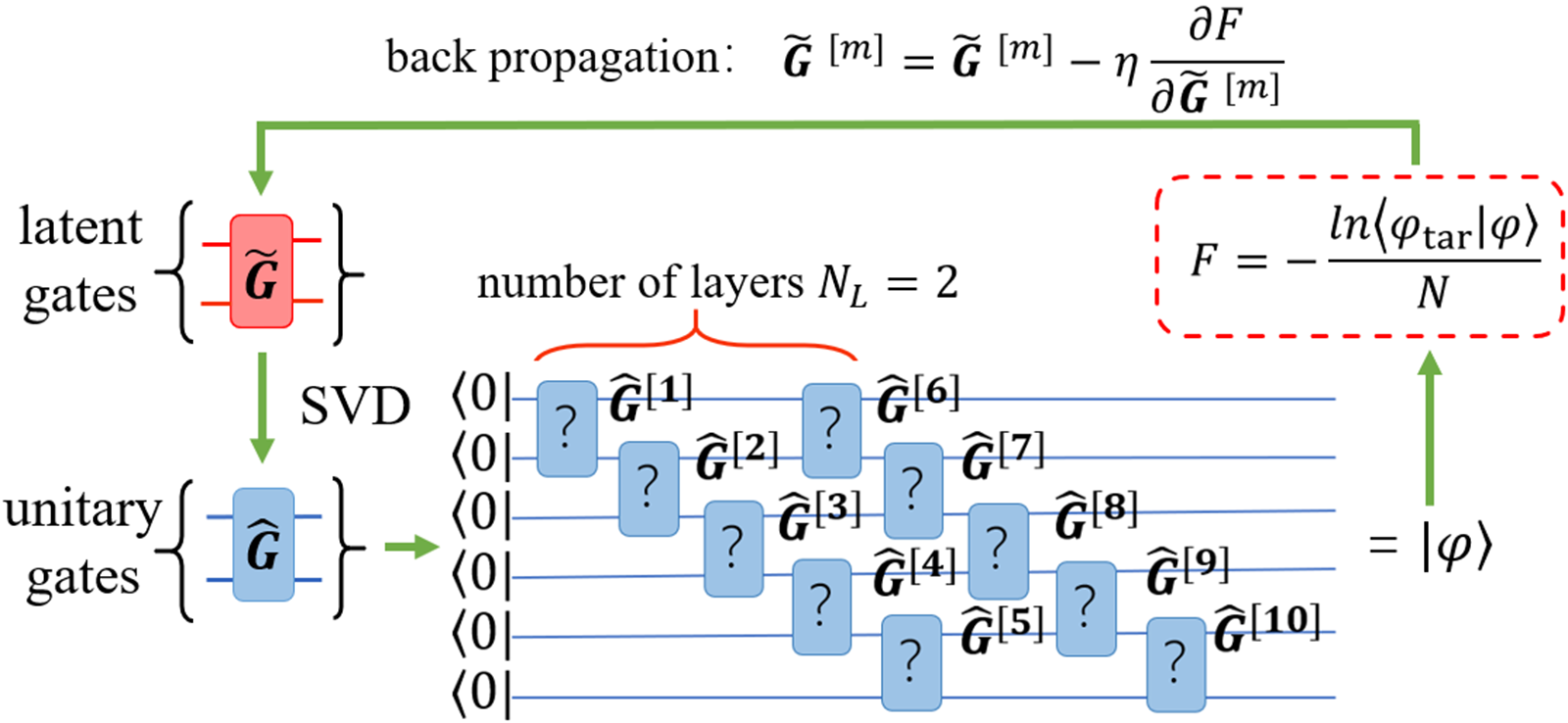}
	\caption{(Color online) Illustration of the optimization process in the automatically differentiable quantum circuit.}
	\label{fig-demon}
\end{figure}


\textit{Automatically differentiable quantum circuit.---} Considering the preparation of a target state $|\psi_{\text{tar}}\rangle$ of $N$ qubits, the task is to find the unitary transformation $\hat{U}$ that maps the initial state $|\psi_0\rangle$ to $|\psi_{\text{tar}}\rangle$ as $|\psi_{\text{tar}}\rangle = \hat{U} |\psi_0\rangle$. Normally, we can take $|\psi_0\rangle$ as a product state, e.g., $|\psi_0\rangle = \prod_{\otimes n=1}^{N} |0_n \rangle$. To realize the unitary transformation on a quantum computer, $\hat{U}$ is normally taken to be a quantum circuit formed by two-qubit gates denoted as $\{\hat{G}\}$. The complexity of the circuit can be characterized by its depth and the number of gates therein. To characterize the accuracy of the state preparation, we choose the negative logarithmic fidelity $F$ per site defined as
\begin{equation}\label{eq-NLF}
F = -\frac{1}{N} \ln |\langle \psi_{\text{tar}}| \hat{U}| \psi_0 \rangle|.
\end{equation} 
For an ideal preparation with the evolved state $|\psi\rangle = \hat{U}|\psi_0\rangle = e^{i\alpha}|\psi_{\text{tar}}\rangle$ where $\alpha$ is a global phase, we have $F=0$.

Our goal is to find the gates $\{\hat{G}\}$ that minimizes $F$ as the loss function, which can be done by updating the gates towards the opposite direction of their gradients. To satisfy the unitary conditions of the quantum gates $\{\hat{G}\}$, we introduce the \text{latent gates} $\{\tilde{G}\}$ that are imposed with no constraints. Each latent gate, say the $m$-th one $\tilde{G}^{[m]}$, gives a quantum gate by $\hat{G}^{[m]} = UV^{\dagger}$ using the singular value decomposition (SVD) $\tilde{G}^{[m]} = U \Lambda V^{\dagger} $. The idea is to project $\tilde{G}^{[m]}$ to the unitary gate that has the largest similarity to it, where $\text{Tr}(\tilde{G}^{[m]} G^{[m]})$ is maximized under the unitary constraint on $\hat{G}^{[m]}$. Such a trick has been used in the entanglement renormalization algorithm~\cite{PhysRevLett.99.220405} and the tree TN machine learning~\cite{LRWP+17MLTN}. The latent gates $\{\tilde{G}\}$ are updated as
\begin{equation}\label{eq-BP}
	\tilde{G}^{[m]} \leftarrow \tilde{G}^{[m]} - \eta \frac{\partial F}{\partial \tilde{G}^{[m]}},
\end{equation} 
where $\eta$ is the learning rate. In practice, $\{\tilde{G}\}$ are given by the automatically differentiable tensors, and their gradients $\frac{\partial F}{\partial \tilde{G}}$ are obtained by the BP method ~\cite{PRXDPTNwanglei2019}. We choose the optimizer Adam and set a learning rate schedule to control $\eta$ during the optimization ~\cite{KB15Adam}. The whole process is illustrated in Fig. \ref{fig-demon}. We choose the stair-like circuit, without losing generality.

One main factor of determining the complexity of our ADQC approach is the number of automatically differentiable parameters, for which we need to compute and save their computational graphs in the BP process. Therefore, we update the gates layer by layer to improve efficiency. To construct a circuit with $N_L$ layers, we start from the first layer (denoted as $\hat{U}^{[1]}$) and optimize the corresponding latent gates by minimizing the NLF $F_1 = -\frac{1}{N} \ln |\langle \psi_{\text{tar}}| \hat{U}^{[1]}|\psi_0  \rangle|$. To increase the number of layers to $(n_L+1)$, for instance, we initialize the latent gates for the first $n_L$ layers by the results from the previous optimizations, and initialize the latent gates in the $(n_L+1)$-th layer as the identities perturbed by small random matrices. Then in each epoch with $(n_L+1)$ layers, we optimize all gates layer by layer by minimizing $F_{n_L} = -\frac{1}{N} \ln |\langle \psi_{\text{tar}}| \prod_{\tilde{n}=n_L+1}^{1} \hat{U}^{[\tilde{n}]} |\psi_0  \rangle|$. After $F_{n_L}$ converges, we add one more layer to the ADQC ($n_L \leftarrow n_L +1$) and repeat the above process, until we have $N_L$ layers in the ADQC. In this way, we only need to deal with the computational graphs associated to the latent gates in one layer in each BP calculation.

The computational complexity is also determined by the simulation of the evolved state. To consider large $N$, we assume $|\psi_{\text{tar}}\rangle$ to be in the form of MPS
\begin{equation}\label{eq-MPS}
	|\psi_{\text{tar}}\rangle = \sum_{a_1 \cdots a_{N-1}} \sum_{s_1 \cdots s_N} A^{[1]}_{s_1 a_1} A^{[2]}_{s_2 a_1 a_{2}} \cdots A^{[N]}_{s_N a_{N-1}} \prod_{\otimes n=1}^N|s_n\rangle,
\end{equation} 
where $|s_n\rangle$ ($s_n=0,1$) form an orthonormal basis of the $n$-th qubit. We set $\dim(a_i) = \chi$ as the virtual bond dimension. The parameter complexity of a $N$-qubit state is reduced from $O(2^N)$ to $O(N\chi^2)$. As any state can be represented as a MPS with sufficiently large $\chi$, assuming $|\psi_{\text{tar}}\rangle$ as a MPS would not harm the arbitrariness of our approach.


\textit{Benchmark results.---} To benchmark ADQC, we choose the target states as the ground states of Heisenberg and XY chains as examples, whose Hamiltonians read
\begin{eqnarray}\label{eq-H}
\hat{H}_{\text{Heisenberg}} &=& \sum_{n=1}^{N-1} (\hat{S}^x_{n} \hat{S}^x_{n+1} + \hat{S}^y_{n} \hat{S}^y_{n+1} + \hat{S}^z_{n} \hat{S}^z_{n+1}), \\
\hat{H}_{\text{XY}} &=& \sum_{n=1}^{N-1} (\hat{S}^x_{n} \hat{S}^x_{n+1} + \hat{S}^y_{n} \hat{S}^y_{n+1}),
\end{eqnarray} 
with $\hat{S}^{\alpha}_{n}$ ($\alpha = x,y,z$) the spin-$\frac{1}{2}$ operators for the $n$-th qubit. 

\begin{figure}[tbp]
	\centering
	\includegraphics[angle=0,width=1\linewidth]{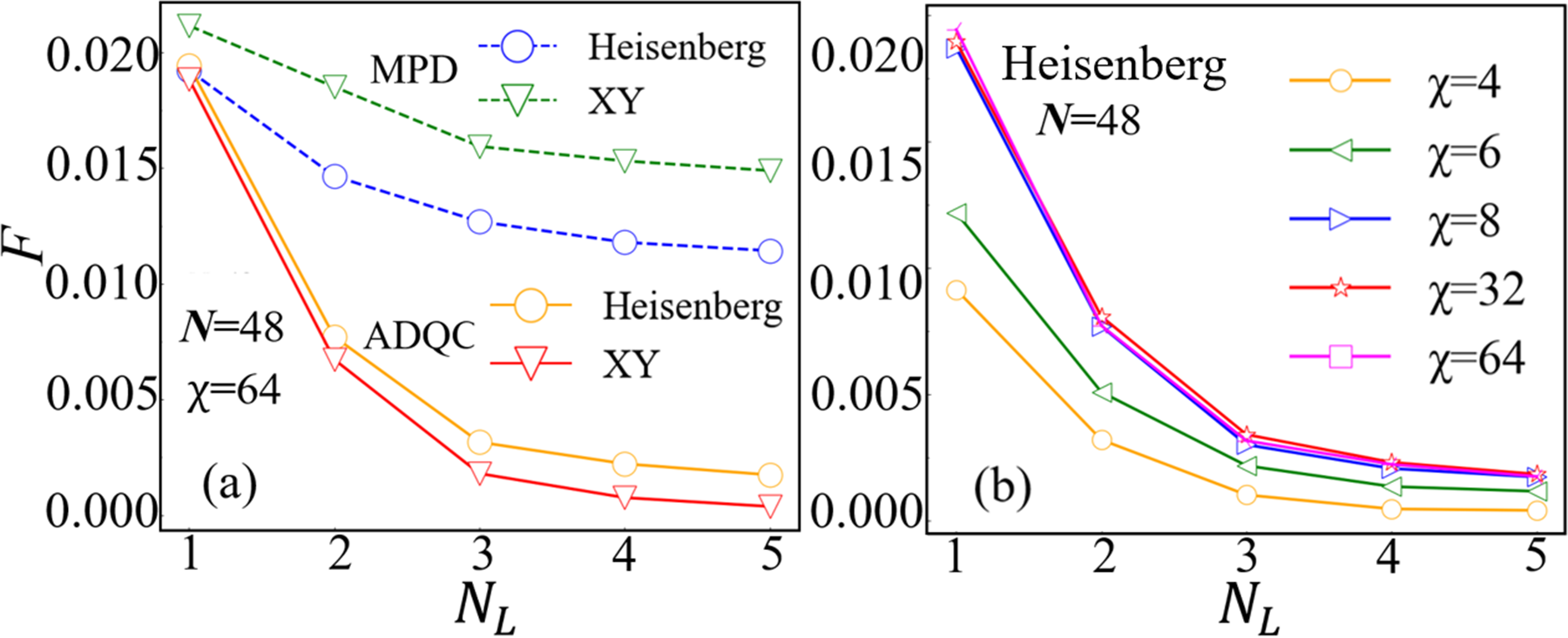}
	\caption{(Color online) Negative logarithmic fidelity $F$ [Eq. (\ref{eq-NLF})] with different number of layers $N_L$ in the ADQC. In (a), we take the number of qubits $N=48$ and the virtual dimensional of the target MPS $\chi=64$ to represent the ground states of the XY and Heisenberg chains. Our results are compared with those from the circuits obtained by the matrix product disentangler (MPD) method {\cite{PhysRevA.101.032310}}. In (b) we show the $F$ by ADQC for preparing the ground-state MPS's of the Heisenberg chain with $\chi=4 \sim 64$.}
	\label{fig-NL}
\end{figure}

We compare our method with the matrix product disentangler (MPD) approach ~\cite{PhysRevA.101.032310}, where the quantum circuit is constructed from the conjugate transpose of the matrix product operator that disentangles the target state. There is no variational process in the MPD method. We take the number of qubits $N=48$, for which it is unlikely to process the full coefficients of the states. As expected, the NLF $F$ decreases when the number of layers in the circuit $N_L$ increases for both ADQC and MPD, as shown in Fig.~\ref{fig-NL}~(a). Our ADQC exhibits lower $F$, particularly for $N_L \geq 2$ thanks to the iterative optimizations in ADQC. Note in MPD, the gates in the $n_L$-th layer are constructed only in order to disentangle $|\psi_{n_L-1} \rangle = \prod_{\tilde{n}=n_L-1}^1 \hat{U}^{[\tilde{n}]} |\psi_{0}\rangle$. In other words, these gates are not optimized by considering the gates in the last $(N_L-n_L)$ layers.

Fig.~\ref{fig-NL}~(b) shows the $F$ by taking different virtual bond dimensions $\chi$ in the MPS that approximates the ground state of the Heisenberg chain. In principle, a quantum circuit with $N_L$ layers can prepare a MPS with the virtual bond dimension at most $\chi = 4^{N_L}$. For $\chi \ll 4^{N_L}$, one may observe obvious rise of $F$ when $\chi$ increases.

The number of parameters in a stair-like ADQC satisfies $\#(\text{ADQC}) = 16(N-1)N_L$, and that of a MPS is given by the number of tensor elements in Eq. (\ref{eq-MPS}) satisfying $\#(\text{MPS}) = 4\chi + 2(N-2)\chi^2$. Our results show that the parameter complexity of a MPS can be significantly compressed by writing the state as the evolution with ADQC, where we have the compression ratio
\begin{equation}\label{eq-r}
r = \frac{\#(\text{ADQC})}{\#(\text{MPS})} \simeq N_L r_0,
\end{equation} 
with $r_0 \sim O(10^{-3})$ for $\chi \leq 32$.

\begin{figure}[tbp]
	\centering
	\includegraphics[angle=0,width=1\linewidth]{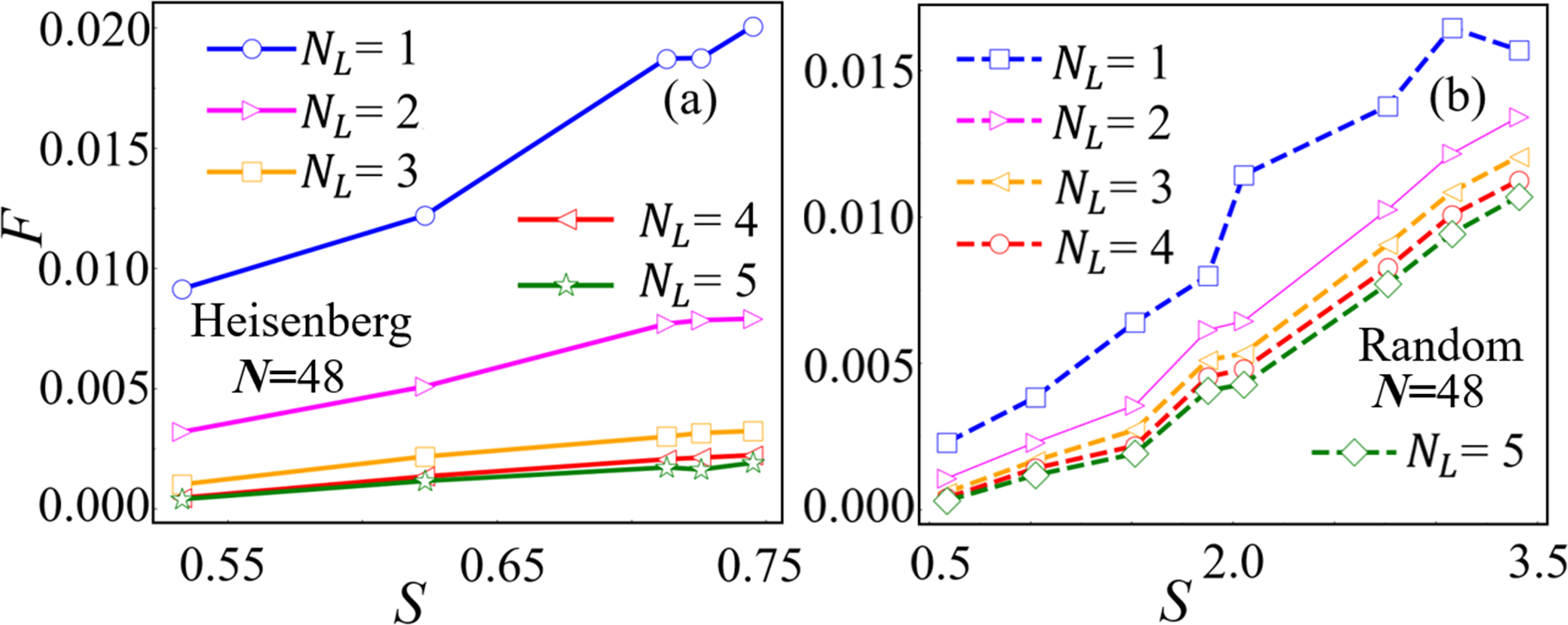}
	\caption{(Color online) Negative logarithmic fidelity $F$ [Eq. (\ref{eq-NLF})] by ADQC for preparing the states with different entanglement entropy $S$ measured in the middle of the system. In (a), we fix the total number of qubits $N=48$. Different values of $S$ are reached by varying $\chi$ of the MPS to approximate the ground state of the Heisenberg chain. In (b), we randomly take the elements of the tensors in the MPS's with different $\chi$ to obtain the target states with different values of $S$.}
	\label{fig-S}
\end{figure}

Essentially, the difficulty of preparing a target state is determined by its entanglement. The upper bond of the bipartite entanglement entropy $S$ of the state prepared by a $N_L$-layer quantum satisfies $S \leq N_L\ln4$. For the Heisenberg and XY chains, the ground states are gapless. According to the conformal field theory ~\cite{moore_classical_1989}, when $N$ is sufficiently large, the entanglement entropy of the ground-state MPS increases logarithmically with $\chi$ as $S =- \frac{c}{3} \ln \chi$ with the central charge $c=1$ for the Heisenberg chain ~\cite{PRLEntanglementQCP2003,PRBTalScalingEntanglement2008,PhysRevLett.102.255701}. Fig.~\ref{fig-S}~(a) demonstrates the $F$ for preparing the target states that possess different entanglement entropies $S$. Note $S$ is measured in the middle of the system by cutting it to two equal halves. Different values of $S$ are obtained by varying the virtual bond dimension $\chi$. The $F$ increases with $S$ as expected.

To further demonstrate the scaling behavior of $F$ against $S$, we take the random MPS's as the target states. In detail, all the tensor elements in the MPS are taken as random numbers, and the MPS's are normalized to satisfy $\langle \psi_{\text{tar}}| \psi_{\text{tar}} \rangle=1$. By taking different random numbers as the tensor elements with different $\chi$, the entanglement entropies of the random MPS's vary from $S\simeq 0.5$ to $3.5$ approximately. Compared with the ground-state MPS's of the quantum spin chains, a random MPS with similar $\chi$ may possess larger $S$, since the $S$ of the ground states should satisfy the 1D area law with a strict upper bound~\cite{PRBTalScalingEntanglement2008}. With a same $N_L$, we obtain similar $F$ for the random MPS's even when $S$ is much larger than that of the ground-state MPS's, particularly for small $N_L$'s. For instance, we have $F\sim O(10^{-1})$ with $N_L = 2$, while the entanglement entropy is $S \simeq 0.75$ for the ground-state MPS of the Heisenberg chain and $S \simeq 3.5$ for the random MPS. These results imply that the difficulty for state preparation should be affected by not only the amount of the entanglement but also its non-trivial structure, such as those described by 1D CFT.

\begin{figure}[tbp]
	\centering
	\includegraphics[angle=0,width=1\linewidth]{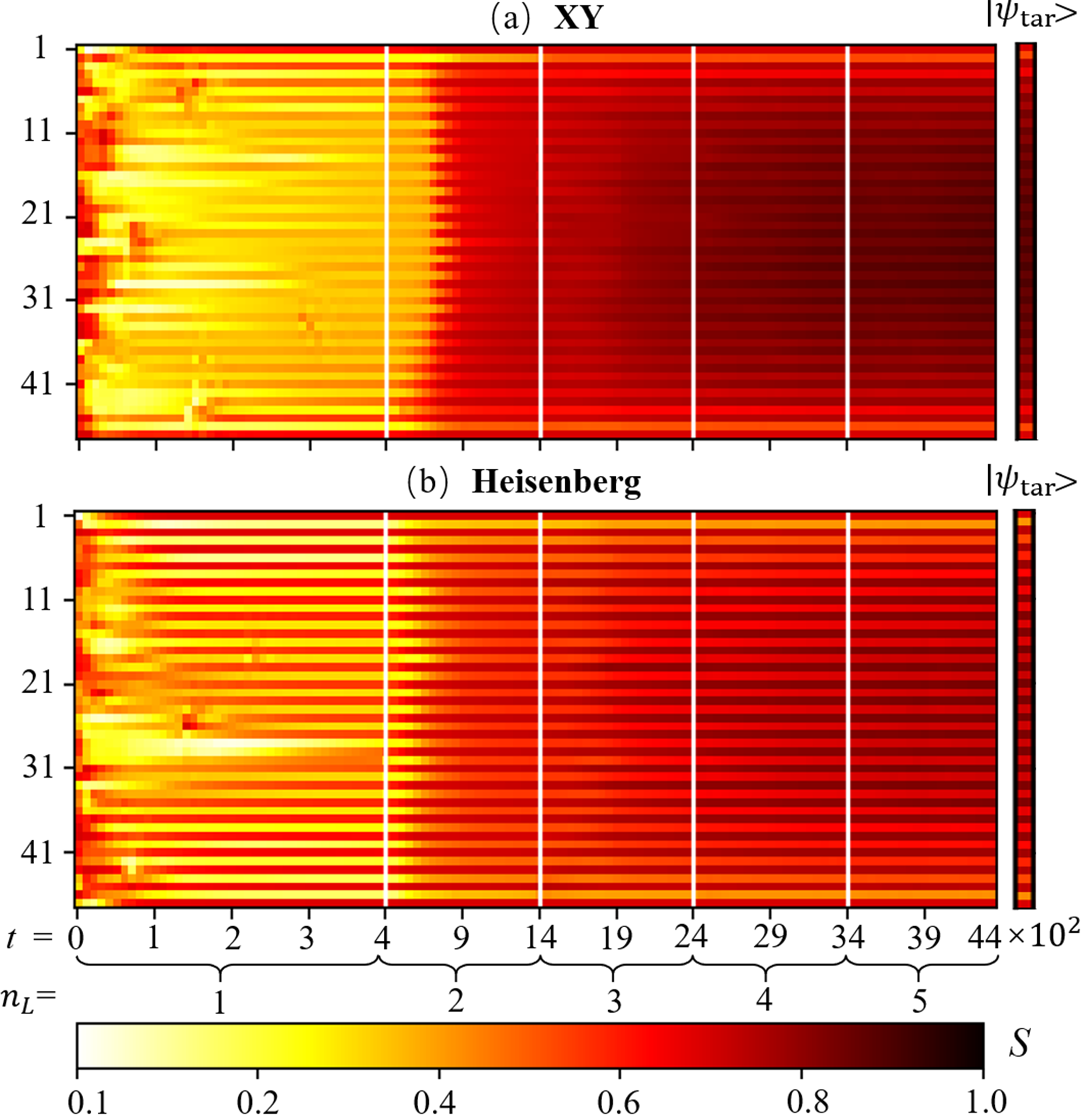}
	\includegraphics[angle=0,width=1\linewidth]{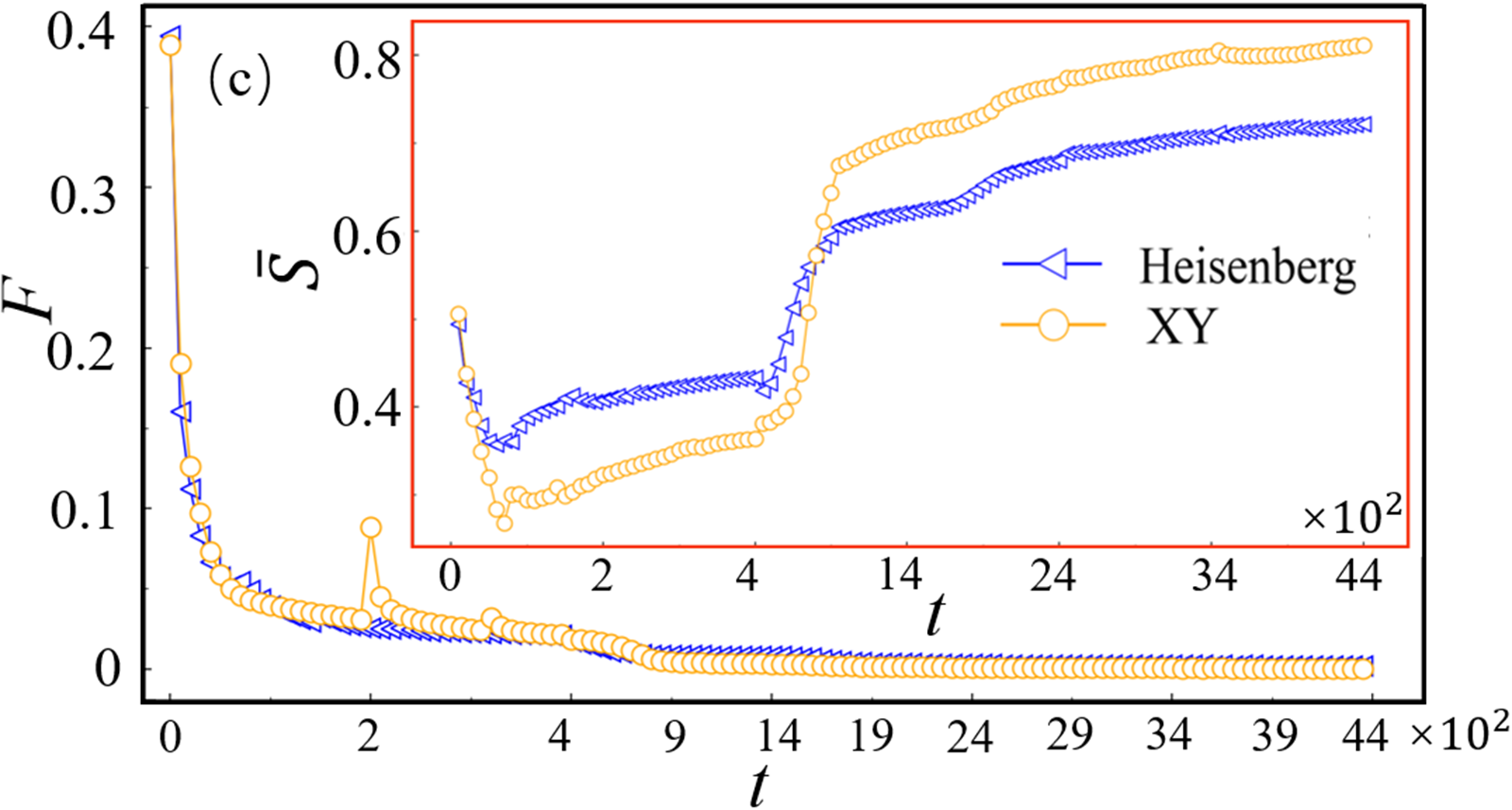}
	\caption{(Color online) (a) and (b) show how the bipartite entanglement entropies $S_n$ by measuring between the $n$-th and $(n+1)$-th qubits vary with the number of epochs $t$. We take the target state $|\psi_{\text{tar}} \rangle$ as the ground-state MPS of the Heisenberg and XY chains, respectively. At different epochs, the ADQC contains different numbers of layers $n_L$. The $S_n$ of $|\psi_{\text{tar}} \rangle$ are given on the right side. In (c) we show the $F$ and the average entanglement entropy $\bar{S}$ (in the inset) for different $t$.}
	\label{fig-hot1}
\end{figure}

To illustrate how the evolved state approaches to the target state $| \psi_{\text{tar}} \rangle$ during the optimization, we calculate the bipartite entanglement entropies $S_n$ by measuring between the $n$-th and $(n+1)$-th qubits. Taking the ground-state MPS's of the XY and Heisenberg chains as $| \psi_{\text{tar}} \rangle$, Fig.~\ref{fig-hot1} (a) and (b), respectively, illustrate the $S_n$ of the evolved states for different number epochs $t$. We set the the total number of layers as $N_L=5$. The $S_n$ of $| \psi_{\text{tar}} \rangle$ are given on the right side of the figures.

In the first $t=400$ epochs, there are only $n_L=1$ layer of gates in the ADQC. As the latent gates are initialized randomly, the unitary gates may be far away from the identity. Thus the $S_n$ for small $t$'s is not small but distributed randomly. For about $t>200$, the evolved state $\hat{U}^{[1]}|\psi_0\rangle$ manages to form the staggered patten in the distribution of $S_n$. During the optimization of the gates with one layer, the $F$ drops fast and converges for about $t=400$, as shown in Fig.~\ref{fig-hot1} (c). The inset shows the average $S_n$ defined as
\begin{equation}\label{eq-OEE}
\bar{S} = \frac{1}{N-1}\sum_{n=1}^{N-1} S_{n},
\end{equation} 
which we use to characterize the strength of entanglement. For $t=1$, the $\bar{S}$ of the evolved state is not small, again due to the random initialization of the latent gates in the first layer. As $F$ decreases, the optimization firstly corrects the false entanglement and then produces the right entanglement structure in the evolved state. This process is indicated by a fast drop then a steady rise of $\bar{S}$ for both the Heisenberg and XY chains.

By adding the second layer to the circuit (in the epochs $400< t \leq 1400$), the optimization mainly enhances the entanglement based on the staggered pattern. We observe a fast increase of $\bar{S}$ for about $400 < t < 1000$. The NLF converges with $n_L=2$ layers around $t=1400$. By adding more layers to the ADQC, the NLF further decreases smoothly in the optimization process. The distribution of $S_{n}$ further approaches to that of the target state. 


\textit{Summary.---} In this work, we propose the automatically differentiable quantum circuit (ADQC) approach, which can derive the circuit to prepare any quantum states including those with large numbers of qubits. The idea is to introduce automatically differentiable latent gates as the variational parameters of ADQC, whose decompositions determine the unitary gates in the circuit. With $N_L \sim O(1)$ layers, high fidelities are obtained preparing the random MPS's and the ground states of quantum spin chains that contain, e.g., $N=48$ qubits. The ADQC possesses significantly lower parameter complexity than the target MPS prepared by the ADQC. Our method can be applied to prepare any states desired in the tasks of quantum computations. Our work paves the way to automatic construction of large-scale quantum circuits by bringing in the machine learning methods.

\textit{Acknowledgment.---} PFZ is thankful to Ding-Zu Wang, Wei-Ming Li, Pei Shi, and Xiao-Han Wang  for stimulating discussions. This work was supported by NSFC (Grant No. 12004266 and No. 11834014), Beijing Natural Science Foundation (No. 1192005 and No. Z180013), Foundation of Beijing Education Committees (No. KM202010028013), and the key research project of Academy for Multidisciplinary Studies, Capital Normal University.


\begin{thebibliography}{39}%
\makeatletter
\providecommand \@ifxundefined [1]{%
 \@ifx{#1\undefined}
}%
\providecommand \@ifnum [1]{%
 \ifnum #1\expandafter \@firstoftwo
 \else \expandafter \@secondoftwo
 \fi
}%
\providecommand \@ifx [1]{%
 \ifx #1\expandafter \@firstoftwo
 \else \expandafter \@secondoftwo
 \fi
}%
\providecommand \natexlab [1]{#1}%
\providecommand \enquote  [1]{``#1''}%
\providecommand \bibnamefont  [1]{#1}%
\providecommand \bibfnamefont [1]{#1}%
\providecommand \citenamefont [1]{#1}%
\providecommand \href@noop [0]{\@secondoftwo}%
\providecommand \href [0]{\begingroup \@sanitize@url \@href}%
\providecommand \@href[1]{\@@startlink{#1}\@@href}%
\providecommand \@@href[1]{\endgroup#1\@@endlink}%
\providecommand \@sanitize@url [0]{\catcode `\\12\catcode `\$12\catcode
  `\&12\catcode `\#12\catcode `\^12\catcode `\_12\catcode `\%12\relax}%
\providecommand \@@startlink[1]{}%
\providecommand \@@endlink[0]{}%
\providecommand \url  [0]{\begingroup\@sanitize@url \@url }%
\providecommand \@url [1]{\endgroup\@href {#1}{\urlprefix }}%
\providecommand \urlprefix  [0]{URL }%
\providecommand \Eprint [0]{\href }%
\providecommand \doibase [0]{https://doi.org/}%
\providecommand \selectlanguage [0]{\@gobble}%
\providecommand \bibinfo  [0]{\@secondoftwo}%
\providecommand \bibfield  [0]{\@secondoftwo}%
\providecommand \translation [1]{[#1]}%
\providecommand \BibitemOpen [0]{}%
\providecommand \bibitemStop [0]{}%
\providecommand \bibitemNoStop [0]{.\EOS\space}%
\providecommand \EOS [0]{\spacefactor3000\relax}%
\providecommand \BibitemShut  [1]{\csname bibitem#1\endcsname}%
\let\auto@bib@innerbib\@empty
\bibitem [{\citenamefont {{Nielsen}}\ \emph {et~al.}(2002)\citenamefont
  {{Nielsen}}, \citenamefont {{Chuang}},\ and\ \citenamefont
  {{Grover}}}]{AJPNMAQCQI2002}%
  \BibitemOpen
  \bibfield  {author} {\bibinfo {author} {\bibfnamefont {M.~A.}\ \bibnamefont
  {{Nielsen}}}, \bibinfo {author} {\bibfnamefont {I.}~\bibnamefont
  {{Chuang}}},\ and\ \bibinfo {author} {\bibfnamefont {L.~K.}\ \bibnamefont
  {{Grover}}},\ }\bibfield  {title} {\bibinfo {title} {{Quantum Computation and
  Quantum Information}},\ }\href {https://doi.org/10.1119/1.1463744} {\bibfield
   {journal} {\bibinfo  {journal} {American Journal of Physics}\ }\textbf
  {\bibinfo {volume} {70}},\ \bibinfo {pages} {558} (\bibinfo {year}
  {2002})}\BibitemShut {NoStop}%
\bibitem [{\citenamefont {Bennett}\ and\ \citenamefont
  {Wiesner}(1992)}]{PhysRevLett.69.2881}%
  \BibitemOpen
  \bibfield  {author} {\bibinfo {author} {\bibfnamefont {C.~H.}\ \bibnamefont
  {Bennett}}\ and\ \bibinfo {author} {\bibfnamefont {S.~J.}\ \bibnamefont
  {Wiesner}},\ }\bibfield  {title} {\bibinfo {title} {Communication via one-
  and two-particle operators on einstein-podolsky-rosen states},\ }\href
  {https://doi.org/10.1103/PhysRevLett.69.2881} {\bibfield  {journal} {\bibinfo
   {journal} {Phys. Rev. Lett.}\ }\textbf {\bibinfo {volume} {69}},\ \bibinfo
  {pages} {2881} (\bibinfo {year} {1992})}\BibitemShut {NoStop}%
\bibitem [{\citenamefont {Mattle}\ \emph {et~al.}(1996)\citenamefont {Mattle},
  \citenamefont {Weinfurter}, \citenamefont {Kwiat},\ and\ \citenamefont
  {Zeilinger}}]{PRLMKWHCodingQC1996}%
  \BibitemOpen
  \bibfield  {author} {\bibinfo {author} {\bibfnamefont {K.}~\bibnamefont
  {Mattle}}, \bibinfo {author} {\bibfnamefont {H.}~\bibnamefont {Weinfurter}},
  \bibinfo {author} {\bibfnamefont {P.~G.}\ \bibnamefont {Kwiat}},\ and\
  \bibinfo {author} {\bibfnamefont {A.}~\bibnamefont {Zeilinger}},\ }\bibfield
  {title} {\bibinfo {title} {Dense coding in experimental quantum
  communication},\ }\href {https://doi.org/10.1103/PhysRevLett.76.4656}
  {\bibfield  {journal} {\bibinfo  {journal} {Phys. Rev. Lett.}\ }\textbf
  {\bibinfo {volume} {76}},\ \bibinfo {pages} {4656} (\bibinfo {year}
  {1996})}\BibitemShut {NoStop}%
\bibitem [{\citenamefont {Briegel}\ \emph {et~al.}(2009)\citenamefont
  {Briegel}, \citenamefont {Browne}, \citenamefont {Dür}, \citenamefont
  {Raussendorf},\ and\ \citenamefont {Van~den
  Nest}}]{NatureMeasurementbased2009}%
  \BibitemOpen
  \bibfield  {author} {\bibinfo {author} {\bibfnamefont {H.~J.}\ \bibnamefont
  {Briegel}}, \bibinfo {author} {\bibfnamefont {D.~E.}\ \bibnamefont {Browne}},
  \bibinfo {author} {\bibfnamefont {W.}~\bibnamefont {Dür}}, \bibinfo {author}
  {\bibfnamefont {R.}~\bibnamefont {Raussendorf}},\ and\ \bibinfo {author}
  {\bibfnamefont {M.}~\bibnamefont {Van~den Nest}},\ }\bibfield  {title}
  {\bibinfo {title} {Measurement-based quantum computation},\ }\href
  {https://doi.org/10.1038/nphys1157} {\bibfield  {journal} {\bibinfo
  {journal} {Nature Physics}\ }\textbf {\bibinfo {volume} {5}},\ \bibinfo
  {pages} {19} (\bibinfo {year} {2009})}\BibitemShut {NoStop}%
\bibitem [{\citenamefont {Bennett}\ \emph {et~al.}(2001)\citenamefont
  {Bennett}, \citenamefont {DiVincenzo}, \citenamefont {Shor}, \citenamefont
  {Smolin}, \citenamefont {Terhal},\ and\ \citenamefont
  {Wootters}}]{PRLBCHRsatePreparation2001}%
  \BibitemOpen
  \bibfield  {author} {\bibinfo {author} {\bibfnamefont {C.~H.}\ \bibnamefont
  {Bennett}}, \bibinfo {author} {\bibfnamefont {D.~P.}\ \bibnamefont
  {DiVincenzo}}, \bibinfo {author} {\bibfnamefont {P.~W.}\ \bibnamefont
  {Shor}}, \bibinfo {author} {\bibfnamefont {J.~A.}\ \bibnamefont {Smolin}},
  \bibinfo {author} {\bibfnamefont {B.~M.}\ \bibnamefont {Terhal}},\ and\
  \bibinfo {author} {\bibfnamefont {W.~K.}\ \bibnamefont {Wootters}},\
  }\bibfield  {title} {\bibinfo {title} {Remote state preparation},\ }\href
  {https://doi.org/10.1103/PhysRevLett.87.077902} {\bibfield  {journal}
  {\bibinfo  {journal} {Phys. Rev. Lett.}\ }\textbf {\bibinfo {volume} {87}},\
  \bibinfo {pages} {077902} (\bibinfo {year} {2001})}\BibitemShut {NoStop}%
\bibitem [{\citenamefont {Resch}\ \emph {et~al.}(2002)\citenamefont {Resch},
  \citenamefont {Lundeen},\ and\ \citenamefont
  {Steinberg}}]{PRLRKJPreparationstate2002}%
  \BibitemOpen
  \bibfield  {author} {\bibinfo {author} {\bibfnamefont {K.~J.}\ \bibnamefont
  {Resch}}, \bibinfo {author} {\bibfnamefont {J.~S.}\ \bibnamefont {Lundeen}},\
  and\ \bibinfo {author} {\bibfnamefont {A.~M.}\ \bibnamefont {Steinberg}},\
  }\bibfield  {title} {\bibinfo {title} {Quantum state preparation and
  conditional coherence},\ }\href
  {https://doi.org/10.1103/PhysRevLett.88.113601} {\bibfield  {journal}
  {\bibinfo  {journal} {Phys. Rev. Lett.}\ }\textbf {\bibinfo {volume} {88}},\
  \bibinfo {pages} {113601} (\bibinfo {year} {2002})}\BibitemShut {NoStop}%
\bibitem [{\citenamefont {Plesch}\ and\ \citenamefont {Časlav
  Brukner}(2011)}]{PRAQPMBPreparationstate2011}%
  \BibitemOpen
  \bibfield  {author} {\bibinfo {author} {\bibfnamefont {M.}~\bibnamefont
  {Plesch}}\ and\ \bibinfo {author} {\bibnamefont {Časlav Brukner}},\
  }\bibfield  {title} {\bibinfo {title} {Quantum-state preparation with
  universal gate decompositions},\ }\href
  {https://doi.org/10.1103/PhysRevA.83.032302} {\bibfield  {journal} {\bibinfo
  {journal} {Phys. Rev. A}\ }\textbf {\bibinfo {volume} {83}},\ \bibinfo
  {pages} {032302} (\bibinfo {year} {2011})}\BibitemShut {NoStop}%
\bibitem [{\citenamefont {Ran}(2020)}]{PhysRevA.101.032310}%
  \BibitemOpen
  \bibfield  {author} {\bibinfo {author} {\bibfnamefont {S.-J.}\ \bibnamefont
  {Ran}},\ }\bibfield  {title} {\bibinfo {title} {Encoding of matrix product
  states into quantum circuits of one- and two-qubit gates},\ }\href
  {https://doi.org/10.1103/PhysRevA.101.032310} {\bibfield  {journal} {\bibinfo
   {journal} {Phys. Rev. A}\ }\textbf {\bibinfo {volume} {101}},\ \bibinfo
  {pages} {032310} (\bibinfo {year} {2020})}\BibitemShut {NoStop}%
\bibitem [{\citenamefont {Araujo}\ \emph {et~al.}(2021)\citenamefont {Araujo},
  \citenamefont {Park}, \citenamefont {Petruccione},\ and\ \citenamefont
  {da~Silva}}]{araujo_divide-and-conquer_2021}%
  \BibitemOpen
  \bibfield  {author} {\bibinfo {author} {\bibfnamefont {I.~F.}\ \bibnamefont
  {Araujo}}, \bibinfo {author} {\bibfnamefont {D.~K.}\ \bibnamefont {Park}},
  \bibinfo {author} {\bibfnamefont {F.}~\bibnamefont {Petruccione}},\ and\
  \bibinfo {author} {\bibfnamefont {A.~J.}\ \bibnamefont {da~Silva}},\
  }\bibfield  {title} {\bibinfo {title} {A divide-and-conquer algorithm for
  quantum state preparation},\ }\href
  {https://doi.org/10.1038/s41598-021-85474-1} {\bibfield  {journal} {\bibinfo
  {journal} {Scientific Reports}\ }\textbf {\bibinfo {volume} {11}},\ \bibinfo
  {pages} {6329} (\bibinfo {year} {2021})}\BibitemShut {NoStop}%
\bibitem [{\citenamefont {Makhlin}\ \emph {et~al.}(2001)\citenamefont
  {Makhlin}, \citenamefont {Sch\"on},\ and\ \citenamefont
  {Shnirman}}]{RevModPhys.73.357}%
  \BibitemOpen
  \bibfield  {author} {\bibinfo {author} {\bibfnamefont {Y.}~\bibnamefont
  {Makhlin}}, \bibinfo {author} {\bibfnamefont {G.}~\bibnamefont {Sch\"on}},\
  and\ \bibinfo {author} {\bibfnamefont {A.}~\bibnamefont {Shnirman}},\
  }\bibfield  {title} {\bibinfo {title} {Quantum-state engineering with
  josephson-junction devices},\ }\href
  {https://doi.org/10.1103/RevModPhys.73.357} {\bibfield  {journal} {\bibinfo
  {journal} {Rev. Mod. Phys.}\ }\textbf {\bibinfo {volume} {73}},\ \bibinfo
  {pages} {357} (\bibinfo {year} {2001})}\BibitemShut {NoStop}%
\bibitem [{\citenamefont {Nirrengarten}\ \emph {et~al.}(2006)\citenamefont
  {Nirrengarten}, \citenamefont {Qarry}, \citenamefont {Roux}, \citenamefont
  {Emmert}, \citenamefont {Nogues}, \citenamefont {Brune}, \citenamefont
  {Raimond},\ and\ \citenamefont {Haroche}}]{PhysRevLett.97.200405}%
  \BibitemOpen
  \bibfield  {author} {\bibinfo {author} {\bibfnamefont {T.}~\bibnamefont
  {Nirrengarten}}, \bibinfo {author} {\bibfnamefont {A.}~\bibnamefont {Qarry}},
  \bibinfo {author} {\bibfnamefont {C.}~\bibnamefont {Roux}}, \bibinfo {author}
  {\bibfnamefont {A.}~\bibnamefont {Emmert}}, \bibinfo {author} {\bibfnamefont
  {G.}~\bibnamefont {Nogues}}, \bibinfo {author} {\bibfnamefont
  {M.}~\bibnamefont {Brune}}, \bibinfo {author} {\bibfnamefont {J.-M.}\
  \bibnamefont {Raimond}},\ and\ \bibinfo {author} {\bibfnamefont
  {S.}~\bibnamefont {Haroche}},\ }\bibfield  {title} {\bibinfo {title}
  {Realization of a superconducting atom chip},\ }\href
  {https://doi.org/10.1103/PhysRevLett.97.200405} {\bibfield  {journal}
  {\bibinfo  {journal} {Phys. Rev. Lett.}\ }\textbf {\bibinfo {volume} {97}},\
  \bibinfo {pages} {200405} (\bibinfo {year} {2006})}\BibitemShut {NoStop}%
\bibitem [{\citenamefont {Chuang}\ and\ \citenamefont
  {Yamamoto}(1995)}]{PhysRevA.52.3489}%
  \BibitemOpen
  \bibfield  {author} {\bibinfo {author} {\bibfnamefont {I.~L.}\ \bibnamefont
  {Chuang}}\ and\ \bibinfo {author} {\bibfnamefont {Y.}~\bibnamefont
  {Yamamoto}},\ }\bibfield  {title} {\bibinfo {title} {Simple quantum
  computer},\ }\href {https://doi.org/10.1103/PhysRevA.52.3489} {\bibfield
  {journal} {\bibinfo  {journal} {Phys. Rev. A}\ }\textbf {\bibinfo {volume}
  {52}},\ \bibinfo {pages} {3489} (\bibinfo {year} {1995})}\BibitemShut
  {NoStop}%
\bibitem [{\citenamefont {Turchette}\ \emph {et~al.}(1995)\citenamefont
  {Turchette}, \citenamefont {Hood}, \citenamefont {Lange}, \citenamefont
  {Mabuchi},\ and\ \citenamefont {Kimble}}]{PhysRevLett.75.4710}%
  \BibitemOpen
  \bibfield  {author} {\bibinfo {author} {\bibfnamefont {Q.~A.}\ \bibnamefont
  {Turchette}}, \bibinfo {author} {\bibfnamefont {C.~J.}\ \bibnamefont {Hood}},
  \bibinfo {author} {\bibfnamefont {W.}~\bibnamefont {Lange}}, \bibinfo
  {author} {\bibfnamefont {H.}~\bibnamefont {Mabuchi}},\ and\ \bibinfo {author}
  {\bibfnamefont {H.~J.}\ \bibnamefont {Kimble}},\ }\bibfield  {title}
  {\bibinfo {title} {Measurement of conditional phase shifts for quantum
  logic},\ }\href {https://doi.org/10.1103/PhysRevLett.75.4710} {\bibfield
  {journal} {\bibinfo  {journal} {Phys. Rev. Lett.}\ }\textbf {\bibinfo
  {volume} {75}},\ \bibinfo {pages} {4710} (\bibinfo {year}
  {1995})}\BibitemShut {NoStop}%
\bibitem [{\citenamefont {Barenco}\ \emph {et~al.}(1995)\citenamefont
  {Barenco}, \citenamefont {Bennett}, \citenamefont {Cleve}, \citenamefont
  {DiVincenzo}, \citenamefont {Margolus}, \citenamefont {Shor}, \citenamefont
  {Sleator}, \citenamefont {Smolin},\ and\ \citenamefont
  {Weinfurter}}]{PRABarennoEGQC1995}%
  \BibitemOpen
  \bibfield  {author} {\bibinfo {author} {\bibfnamefont {A.}~\bibnamefont
  {Barenco}}, \bibinfo {author} {\bibfnamefont {C.~H.}\ \bibnamefont
  {Bennett}}, \bibinfo {author} {\bibfnamefont {R.}~\bibnamefont {Cleve}},
  \bibinfo {author} {\bibfnamefont {D.~P.}\ \bibnamefont {DiVincenzo}},
  \bibinfo {author} {\bibfnamefont {N.}~\bibnamefont {Margolus}}, \bibinfo
  {author} {\bibfnamefont {P.}~\bibnamefont {Shor}}, \bibinfo {author}
  {\bibfnamefont {T.}~\bibnamefont {Sleator}}, \bibinfo {author} {\bibfnamefont
  {J.~A.}\ \bibnamefont {Smolin}},\ and\ \bibinfo {author} {\bibfnamefont
  {H.}~\bibnamefont {Weinfurter}},\ }\bibfield  {title} {\bibinfo {title}
  {Elementary gates for quantum computation},\ }\href
  {https://doi.org/10.1103/PhysRevA.52.3457} {\bibfield  {journal} {\bibinfo
  {journal} {Phys. Rev. A}\ }\textbf {\bibinfo {volume} {52}},\ \bibinfo
  {pages} {3457} (\bibinfo {year} {1995})}\BibitemShut {NoStop}%
\bibitem [{\citenamefont {DiVincenzo}(1995)}]{PRADivtwobitqc1995}%
  \BibitemOpen
  \bibfield  {author} {\bibinfo {author} {\bibfnamefont {D.~P.}\ \bibnamefont
  {DiVincenzo}},\ }\bibfield  {title} {\bibinfo {title} {Two-bit gates are
  universal for quantum computation},\ }\href
  {https://doi.org/10.1103/PhysRevA.51.1015} {\bibfield  {journal} {\bibinfo
  {journal} {Phys. Rev. A}\ }\textbf {\bibinfo {volume} {51}},\ \bibinfo
  {pages} {1015} (\bibinfo {year} {1995})}\BibitemShut {NoStop}%
\bibitem [{\citenamefont {Rudner}\ and\ \citenamefont
  {Levitov}(2009)}]{PhysRevLett.102.065703}%
  \BibitemOpen
  \bibfield  {author} {\bibinfo {author} {\bibfnamefont {M.~S.}\ \bibnamefont
  {Rudner}}\ and\ \bibinfo {author} {\bibfnamefont {L.~S.}\ \bibnamefont
  {Levitov}},\ }\bibfield  {title} {\bibinfo {title} {Topological transition in
  a non-hermitian quantum walk},\ }\href
  {https://doi.org/10.1103/PhysRevLett.102.065703} {\bibfield  {journal}
  {\bibinfo  {journal} {Phys. Rev. Lett.}\ }\textbf {\bibinfo {volume} {102}},\
  \bibinfo {pages} {065703} (\bibinfo {year} {2009})}\BibitemShut {NoStop}%
\bibitem [{\citenamefont {Harrow}\ and\ \citenamefont
  {Montanaro}(2017)}]{HM17QS2017}%
  \BibitemOpen
  \bibfield  {author} {\bibinfo {author} {\bibfnamefont {A.~W.}\ \bibnamefont
  {Harrow}}\ and\ \bibinfo {author} {\bibfnamefont {A.}~\bibnamefont
  {Montanaro}},\ }\bibfield  {title} {\bibinfo {title} {Quantum computational
  supremacy},\ }\href {https://doi.org/10.1038/nature23458} {\bibfield
  {journal} {\bibinfo  {journal} {Nature}\ }\textbf {\bibinfo {volume} {549}},\
  \bibinfo {pages} {203} (\bibinfo {year} {2017})}\BibitemShut {NoStop}%
\bibitem [{\citenamefont {Arute}\ \emph {et~al.}(2019)\citenamefont {Arute},
  \citenamefont {Arya}, \citenamefont {Babbush}, \citenamefont {Bacon},
  \citenamefont {Bardin}, \citenamefont {Barends}, \citenamefont {Biswas},
  \citenamefont {Boixo}, \citenamefont {Brandao}, \citenamefont {Buell} \emph
  {et~al.}}]{AABB+2019GoogleQ}%
  \BibitemOpen
  \bibfield  {author} {\bibinfo {author} {\bibfnamefont {F.}~\bibnamefont
  {Arute}}, \bibinfo {author} {\bibfnamefont {K.}~\bibnamefont {Arya}},
  \bibinfo {author} {\bibfnamefont {R.}~\bibnamefont {Babbush}}, \bibinfo
  {author} {\bibfnamefont {D.}~\bibnamefont {Bacon}}, \bibinfo {author}
  {\bibfnamefont {J.~C.}\ \bibnamefont {Bardin}}, \bibinfo {author}
  {\bibfnamefont {R.}~\bibnamefont {Barends}}, \bibinfo {author} {\bibfnamefont
  {R.}~\bibnamefont {Biswas}}, \bibinfo {author} {\bibfnamefont
  {S.}~\bibnamefont {Boixo}}, \bibinfo {author} {\bibfnamefont {F.~G.}\
  \bibnamefont {Brandao}}, \bibinfo {author} {\bibfnamefont {D.~A.}\
  \bibnamefont {Buell}}, \emph {et~al.},\ }\bibfield  {title} {\bibinfo {title}
  {Quantum supremacy using a programmable superconducting processor},\ }\href
  {https://doi.org/10.1038/s41586-019-1666-5} {\bibfield  {journal} {\bibinfo
  {journal} {Nature}\ }\textbf {\bibinfo {volume} {574}},\ \bibinfo {pages}
  {505} (\bibinfo {year} {2019})}\BibitemShut {NoStop}%
\bibitem [{\citenamefont {Verstraete}\ \emph {et~al.}(2008)\citenamefont
  {Verstraete}, \citenamefont {Murg},\ and\ \citenamefont
  {Cirac}}]{VMC08MPSPEPSRev2008}%
  \BibitemOpen
  \bibfield  {author} {\bibinfo {author} {\bibfnamefont {F.}~\bibnamefont
  {Verstraete}}, \bibinfo {author} {\bibfnamefont {V.}~\bibnamefont {Murg}},\
  and\ \bibinfo {author} {\bibfnamefont {J.~I.}\ \bibnamefont {Cirac}},\
  }\bibfield  {title} {\bibinfo {title} {{Matrix product states, projected
  entangled pair states, and variational renormalization group methods for
  quantum spin systems}},\ }\href {https://doi.org/10.1080/14789940801912366}
  {\bibfield  {journal} {\bibinfo  {journal} {Advances in Physics}\ }\textbf
  {\bibinfo {volume} {57}},\ \bibinfo {pages} {143} (\bibinfo {year}
  {2008})}\BibitemShut {NoStop}%
\bibitem [{\citenamefont {Cirac}\ and\ \citenamefont
  {Verstraete}(2009)}]{CV09TNSRev2009}%
  \BibitemOpen
  \bibfield  {author} {\bibinfo {author} {\bibfnamefont {J.~I.}\ \bibnamefont
  {Cirac}}\ and\ \bibinfo {author} {\bibfnamefont {F.}~\bibnamefont
  {Verstraete}},\ }\bibfield  {title} {\bibinfo {title} {{Renormalization and
  tensor product states in spin chains and lattices}},\ }\href
  {https://doi.org/10.1088/1751-8113/42/50/504004} {\bibfield  {journal}
  {\bibinfo  {journal} {J. Phys. A: Math. Theor.}\ }\textbf {\bibinfo {volume}
  {42}},\ \bibinfo {pages} {504004} (\bibinfo {year} {2009})}\BibitemShut
  {NoStop}%
\bibitem [{\citenamefont {Or\'{u}s}(2014)}]{O14TNSRev2014}%
  \BibitemOpen
  \bibfield  {author} {\bibinfo {author} {\bibfnamefont {R.}~\bibnamefont
  {Or\'{u}s}},\ }\bibfield  {title} {\bibinfo {title} {{A practical
  introduction to tensor networks: Matrix product states and projected
  entangled pair states}},\ }\href {https://doi.org/10.1016/j.aop.2014.06.013}
  {\bibfield  {journal} {\bibinfo  {journal} {Ann. Phys.}\ }\textbf {\bibinfo
  {volume} {349}},\ \bibinfo {pages} {117} (\bibinfo {year}
  {2014})}\BibitemShut {NoStop}%
\bibitem [{\citenamefont {Ran}\ \emph {et~al.}(2020)\citenamefont {Ran},
  \citenamefont {Tirrito}, \citenamefont {Peng}, \citenamefont {Chen},
  \citenamefont {Tagliacozzo}, \citenamefont {Su},\ and\ \citenamefont
  {Lewenstein}}]{RTPC+17TNrev2020}%
  \BibitemOpen
  \bibfield  {author} {\bibinfo {author} {\bibfnamefont {S.-J.}\ \bibnamefont
  {Ran}}, \bibinfo {author} {\bibfnamefont {E.}~\bibnamefont {Tirrito}},
  \bibinfo {author} {\bibfnamefont {C.}~\bibnamefont {Peng}}, \bibinfo {author}
  {\bibfnamefont {X.}~\bibnamefont {Chen}}, \bibinfo {author} {\bibfnamefont
  {L.}~\bibnamefont {Tagliacozzo}}, \bibinfo {author} {\bibfnamefont
  {G.}~\bibnamefont {Su}},\ and\ \bibinfo {author} {\bibfnamefont
  {M.}~\bibnamefont {Lewenstein}},\ }\href
  {https://doi.org/10.1007/978-3-030-34489-4} {\emph {\bibinfo {title} {Tensor
  Network Contractions: Methods and Applications to Quantum Many-Body
  Systems}}}\ (\bibinfo  {publisher} {Springer, Cham},\ \bibinfo {year}
  {2020})\BibitemShut {NoStop}%
\bibitem [{\citenamefont {Or{\'u}s}(2019)}]{O19TNrev2019}%
  \BibitemOpen
  \bibfield  {author} {\bibinfo {author} {\bibfnamefont {R.}~\bibnamefont
  {Or{\'u}s}},\ }\bibfield  {title} {\bibinfo {title} {Tensor networks for
  complex quantum systems},\ }\href {https://doi.org/10.1038/s42254-019-0086-7}
  {\bibfield  {journal} {\bibinfo  {journal} {Nature Reviews Physics}\ }\textbf
  {\bibinfo {volume} {1}},\ \bibinfo {pages} {538} (\bibinfo {year}
  {2019})}\BibitemShut {NoStop}%
\bibitem [{\citenamefont {Srednicki}(1993)}]{S93Sarea1993}%
  \BibitemOpen
  \bibfield  {author} {\bibinfo {author} {\bibfnamefont {M.}~\bibnamefont
  {Srednicki}},\ }\bibfield  {title} {\bibinfo {title} {{Entropy and area}},\
  }\href {https://doi.org/10.1103/PhysRevLett.71.666} {\bibfield  {journal}
  {\bibinfo  {journal} {Phys. Rev. Lett.}\ }\textbf {\bibinfo {volume} {71}},\
  \bibinfo {pages} {666} (\bibinfo {year} {1993})}\BibitemShut {NoStop}%
\bibitem [{\citenamefont {Schuch}\ \emph {et~al.}(2008)\citenamefont {Schuch},
  \citenamefont {Wolf}, \citenamefont {Verstraete},\ and\ \citenamefont
  {Cirac}}]{SWVC08MPSent2008}%
  \BibitemOpen
  \bibfield  {author} {\bibinfo {author} {\bibfnamefont {N.}~\bibnamefont
  {Schuch}}, \bibinfo {author} {\bibfnamefont {M.~M.}\ \bibnamefont {Wolf}},
  \bibinfo {author} {\bibfnamefont {F.}~\bibnamefont {Verstraete}},\ and\
  \bibinfo {author} {\bibfnamefont {J.~I.}\ \bibnamefont {Cirac}},\ }\bibfield
  {title} {\bibinfo {title} {{Entropy Scaling and Simulability by Matrix
  Product States}},\ }\href {https://doi.org/10.1103/PhysRevLett.100.030504}
  {\bibfield  {journal} {\bibinfo  {journal} {Phys. Rev. Lett.}\ }\textbf
  {\bibinfo {volume} {100}},\ \bibinfo {pages} {030504} (\bibinfo {year}
  {2008})}\BibitemShut {NoStop}%
\bibitem [{\citenamefont {Eisert}\ \emph {et~al.}(2010)\citenamefont {Eisert},
  \citenamefont {Cramer},\ and\ \citenamefont {Plenio}}]{ECP10AreaLawRev2010}%
  \BibitemOpen
  \bibfield  {author} {\bibinfo {author} {\bibfnamefont {J.}~\bibnamefont
  {Eisert}}, \bibinfo {author} {\bibfnamefont {M.}~\bibnamefont {Cramer}},\
  and\ \bibinfo {author} {\bibfnamefont {M.~B.}\ \bibnamefont {Plenio}},\
  }\bibfield  {title} {\bibinfo {title} {Colloquium: Area laws for the
  entanglement entropy},\ }\href {https://doi.org/10.1103/RevModPhys.82.277}
  {\bibfield  {journal} {\bibinfo  {journal} {Rev. Mod. Phys.}\ }\textbf
  {\bibinfo {volume} {82}},\ \bibinfo {pages} {277} (\bibinfo {year}
  {2010})}\BibitemShut {NoStop}%
\bibitem [{\citenamefont {Mitarai}\ \emph {et~al.}(2018)\citenamefont
  {Mitarai}, \citenamefont {Negoro}, \citenamefont {Kitagawa},\ and\
  \citenamefont {Fujii}}]{PhysRevA.98.032309}%
  \BibitemOpen
  \bibfield  {author} {\bibinfo {author} {\bibfnamefont {K.}~\bibnamefont
  {Mitarai}}, \bibinfo {author} {\bibfnamefont {M.}~\bibnamefont {Negoro}},
  \bibinfo {author} {\bibfnamefont {M.}~\bibnamefont {Kitagawa}},\ and\
  \bibinfo {author} {\bibfnamefont {K.}~\bibnamefont {Fujii}},\ }\bibfield
  {title} {\bibinfo {title} {Quantum circuit learning},\ }\href
  {https://doi.org/10.1103/PhysRevA.98.032309} {\bibfield  {journal} {\bibinfo
  {journal} {Phys. Rev. A}\ }\textbf {\bibinfo {volume} {98}},\ \bibinfo
  {pages} {032309} (\bibinfo {year} {2018})}\BibitemShut {NoStop}%
\bibitem [{\citenamefont {Liu}\ and\ \citenamefont
  {Wang}(2018)}]{PhysRevA.98.062324}%
  \BibitemOpen
  \bibfield  {author} {\bibinfo {author} {\bibfnamefont {J.-G.}\ \bibnamefont
  {Liu}}\ and\ \bibinfo {author} {\bibfnamefont {L.}~\bibnamefont {Wang}},\
  }\bibfield  {title} {\bibinfo {title} {Differentiable learning of quantum
  circuit born machines},\ }\href {https://doi.org/10.1103/PhysRevA.98.062324}
  {\bibfield  {journal} {\bibinfo  {journal} {Phys. Rev. A}\ }\textbf {\bibinfo
  {volume} {98}},\ \bibinfo {pages} {062324} (\bibinfo {year}
  {2018})}\BibitemShut {NoStop}%
\bibitem [{\citenamefont {Arrazola}\ \emph {et~al.}(2019)\citenamefont
  {Arrazola}, \citenamefont {Bromley}, \citenamefont {Izaac}, \citenamefont
  {Myers}, \citenamefont {Br{\'{a}}dler},\ and\ \citenamefont
  {Killoran}}]{Arrazola_2019}%
  \BibitemOpen
  \bibfield  {author} {\bibinfo {author} {\bibfnamefont {J.~M.}\ \bibnamefont
  {Arrazola}}, \bibinfo {author} {\bibfnamefont {T.~R.}\ \bibnamefont
  {Bromley}}, \bibinfo {author} {\bibfnamefont {J.}~\bibnamefont {Izaac}},
  \bibinfo {author} {\bibfnamefont {C.~R.}\ \bibnamefont {Myers}}, \bibinfo
  {author} {\bibfnamefont {K.}~\bibnamefont {Br{\'{a}}dler}},\ and\ \bibinfo
  {author} {\bibfnamefont {N.}~\bibnamefont {Killoran}},\ }\bibfield  {title}
  {\bibinfo {title} {Machine learning method for state preparation and gate
  synthesis on photonic quantum computers},\ }\href
  {https://doi.org/10.1088/2058-9565/aaf59e} {\bibfield  {journal} {\bibinfo
  {journal} {Quantum Science and Technology}\ }\textbf {\bibinfo {volume}
  {4}},\ \bibinfo {pages} {024004} (\bibinfo {year} {2019})}\BibitemShut
  {NoStop}%
\bibitem [{\citenamefont {Ghosh}\ \emph {et~al.}(2019)\citenamefont {Ghosh},
  \citenamefont {Paterek},\ and\ \citenamefont
  {Liew}}]{PhysRevLett.123.260404}%
  \BibitemOpen
  \bibfield  {author} {\bibinfo {author} {\bibfnamefont {S.}~\bibnamefont
  {Ghosh}}, \bibinfo {author} {\bibfnamefont {T.}~\bibnamefont {Paterek}},\
  and\ \bibinfo {author} {\bibfnamefont {T.~C.~H.}\ \bibnamefont {Liew}},\
  }\bibfield  {title} {\bibinfo {title} {Quantum neuromorphic platform for
  quantum state preparation},\ }\href
  {https://doi.org/10.1103/PhysRevLett.123.260404} {\bibfield  {journal}
  {\bibinfo  {journal} {Phys. Rev. Lett.}\ }\textbf {\bibinfo {volume} {123}},\
  \bibinfo {pages} {260404} (\bibinfo {year} {2019})}\BibitemShut {NoStop}%
\bibitem [{\citenamefont {Liao}\ \emph {et~al.}(2019)\citenamefont {Liao},
  \citenamefont {Liu}, \citenamefont {Wang},\ and\ \citenamefont
  {Xiang}}]{PRXDPTNwanglei2019}%
  \BibitemOpen
  \bibfield  {author} {\bibinfo {author} {\bibfnamefont {H.-J.}\ \bibnamefont
  {Liao}}, \bibinfo {author} {\bibfnamefont {J.-G.}\ \bibnamefont {Liu}},
  \bibinfo {author} {\bibfnamefont {L.}~\bibnamefont {Wang}},\ and\ \bibinfo
  {author} {\bibfnamefont {T.}~\bibnamefont {Xiang}},\ }\bibfield  {title}
  {\bibinfo {title} {Differentiable programming tensor networks},\ }\href
  {https://doi.org/10.1103/PhysRevX.9.031041} {\bibfield  {journal} {\bibinfo
  {journal} {Phys. Rev. X}\ }\textbf {\bibinfo {volume} {9}},\ \bibinfo {pages}
  {031041} (\bibinfo {year} {2019})}\BibitemShut {NoStop}%
\bibitem [{\citenamefont {Chen}\ \emph {et~al.}(2020)\citenamefont {Chen},
  \citenamefont {Gao}, \citenamefont {Guo}, \citenamefont {Liu}, \citenamefont
  {Zhao}, \citenamefont {Liao}, \citenamefont {Wang}, \citenamefont {Xiang},
  \citenamefont {Li},\ and\ \citenamefont {Xie}}]{dTRG2020}%
  \BibitemOpen
  \bibfield  {author} {\bibinfo {author} {\bibfnamefont {B.-B.}\ \bibnamefont
  {Chen}}, \bibinfo {author} {\bibfnamefont {Y.}~\bibnamefont {Gao}}, \bibinfo
  {author} {\bibfnamefont {Y.-B.}\ \bibnamefont {Guo}}, \bibinfo {author}
  {\bibfnamefont {Y.}~\bibnamefont {Liu}}, \bibinfo {author} {\bibfnamefont
  {H.-H.}\ \bibnamefont {Zhao}}, \bibinfo {author} {\bibfnamefont {H.-J.}\
  \bibnamefont {Liao}}, \bibinfo {author} {\bibfnamefont {L.}~\bibnamefont
  {Wang}}, \bibinfo {author} {\bibfnamefont {T.}~\bibnamefont {Xiang}},
  \bibinfo {author} {\bibfnamefont {W.}~\bibnamefont {Li}},\ and\ \bibinfo
  {author} {\bibfnamefont {Z.~Y.}\ \bibnamefont {Xie}},\ }\bibfield  {title}
  {\bibinfo {title} {Automatic differentiation for second renormalization of
  tensor networks},\ }\href {https://doi.org/10.1103/PhysRevB.101.220409}
  {\bibfield  {journal} {\bibinfo  {journal} {Phys. Rev. B}\ }\textbf {\bibinfo
  {volume} {101}},\ \bibinfo {pages} {220409} (\bibinfo {year}
  {2020})}\BibitemShut {NoStop}%
\bibitem [{\citenamefont {Vidal}(2007)}]{PhysRevLett.99.220405}%
  \BibitemOpen
  \bibfield  {author} {\bibinfo {author} {\bibfnamefont {G.}~\bibnamefont
  {Vidal}},\ }\bibfield  {title} {\bibinfo {title} {Entanglement
  renormalization},\ }\href {https://doi.org/10.1103/PhysRevLett.99.220405}
  {\bibfield  {journal} {\bibinfo  {journal} {Phys. Rev. Lett.}\ }\textbf
  {\bibinfo {volume} {99}},\ \bibinfo {pages} {220405} (\bibinfo {year}
  {2007})}\BibitemShut {NoStop}%
\bibitem [{\citenamefont {Liu}\ \emph {et~al.}(2019)\citenamefont {Liu},
  \citenamefont {Ran}, \citenamefont {Wittek}, \citenamefont {Peng},
  \citenamefont {Garc{\'{\i}}a}, \citenamefont {Su},\ and\ \citenamefont
  {Lewenstein}}]{LRWP+17MLTN}%
  \BibitemOpen
  \bibfield  {author} {\bibinfo {author} {\bibfnamefont {D.}~\bibnamefont
  {Liu}}, \bibinfo {author} {\bibfnamefont {S.-J.}\ \bibnamefont {Ran}},
  \bibinfo {author} {\bibfnamefont {P.}~\bibnamefont {Wittek}}, \bibinfo
  {author} {\bibfnamefont {C.}~\bibnamefont {Peng}}, \bibinfo {author}
  {\bibfnamefont {R.~B.}\ \bibnamefont {Garc{\'{\i}}a}}, \bibinfo {author}
  {\bibfnamefont {G.}~\bibnamefont {Su}},\ and\ \bibinfo {author}
  {\bibfnamefont {M.}~\bibnamefont {Lewenstein}},\ }\bibfield  {title}
  {\bibinfo {title} {Machine learning by unitary tensor network of hierarchical
  tree structure},\ }\href {https://doi.org/10.1088/1367-2630/ab31ef}
  {\bibfield  {journal} {\bibinfo  {journal} {New Journal of Physics}\ }\textbf
  {\bibinfo {volume} {21}},\ \bibinfo {pages} {073059} (\bibinfo {year}
  {2019})}\BibitemShut {NoStop}%
\bibitem [{\citenamefont {Kingma}\ and\ \citenamefont {Ba}(2015)}]{KB15Adam}%
  \BibitemOpen
  \bibfield  {author} {\bibinfo {author} {\bibfnamefont {D.~P.}\ \bibnamefont
  {Kingma}}\ and\ \bibinfo {author} {\bibfnamefont {J.}~\bibnamefont {Ba}},\
  }\bibfield  {title} {\bibinfo {title} {Adam: {A} method for stochastic
  optimization},\ }in\ \href {http://arxiv.org/abs/1412.6980} {\emph {\bibinfo
  {booktitle} {3rd International Conference on Learning Representations, {ICLR}
  2015, San Diego, CA, USA, May 7-9, 2015, Conference Track Proceedings}}}\
  (\bibinfo {year} {2015})\BibitemShut {NoStop}%
\bibitem [{\citenamefont {Moore}\ and\ \citenamefont
  {Seiberg}(1989)}]{moore_classical_1989}%
  \BibitemOpen
  \bibfield  {author} {\bibinfo {author} {\bibfnamefont {G.}~\bibnamefont
  {Moore}}\ and\ \bibinfo {author} {\bibfnamefont {N.}~\bibnamefont
  {Seiberg}},\ }\bibfield  {title} {\bibinfo {title} {Classical and quantum
  conformal field theory},\ }\href {https://doi.org/10.1007/BF01238857}
  {\bibfield  {journal} {\bibinfo  {journal} {Communications in Mathematical
  Physics}\ }\textbf {\bibinfo {volume} {123}},\ \bibinfo {pages} {177}
  (\bibinfo {year} {1989})}\BibitemShut {NoStop}%
\bibitem [{\citenamefont {Vidal}\ \emph {et~al.}(2003)\citenamefont {Vidal},
  \citenamefont {Latorre}, \citenamefont {Rico},\ and\ \citenamefont
  {Kitaev}}]{PRLEntanglementQCP2003}%
  \BibitemOpen
  \bibfield  {author} {\bibinfo {author} {\bibfnamefont {G.}~\bibnamefont
  {Vidal}}, \bibinfo {author} {\bibfnamefont {J.~I.}\ \bibnamefont {Latorre}},
  \bibinfo {author} {\bibfnamefont {E.}~\bibnamefont {Rico}},\ and\ \bibinfo
  {author} {\bibfnamefont {A.}~\bibnamefont {Kitaev}},\ }\bibfield  {title}
  {\bibinfo {title} {Entanglement in quantum critical phenomena},\ }\href
  {https://doi.org/10.1103/PhysRevLett.90.227902} {\bibfield  {journal}
  {\bibinfo  {journal} {Phys. Rev. Lett.}\ }\textbf {\bibinfo {volume} {90}},\
  \bibinfo {pages} {227902} (\bibinfo {year} {2003})}\BibitemShut {NoStop}%
\bibitem [{\citenamefont {Tagliacozzo}\ \emph {et~al.}(2008)\citenamefont
  {Tagliacozzo}, \citenamefont {de~Oliveira}, \citenamefont {Iblisdir},\ and\
  \citenamefont {Latorre}}]{PRBTalScalingEntanglement2008}%
  \BibitemOpen
  \bibfield  {author} {\bibinfo {author} {\bibfnamefont {L.}~\bibnamefont
  {Tagliacozzo}}, \bibinfo {author} {\bibfnamefont {T.~R.}\ \bibnamefont
  {de~Oliveira}}, \bibinfo {author} {\bibfnamefont {S.}~\bibnamefont
  {Iblisdir}},\ and\ \bibinfo {author} {\bibfnamefont {J.~I.}\ \bibnamefont
  {Latorre}},\ }\bibfield  {title} {\bibinfo {title} {Scaling of entanglement
  support for matrix product states},\ }\href
  {https://doi.org/10.1103/PhysRevB.78.024410} {\bibfield  {journal} {\bibinfo
  {journal} {Phys. Rev. B}\ }\textbf {\bibinfo {volume} {78}},\ \bibinfo
  {pages} {024410} (\bibinfo {year} {2008})}\BibitemShut {NoStop}%
\bibitem [{\citenamefont {Pollmann}\ \emph {et~al.}(2009)\citenamefont
  {Pollmann}, \citenamefont {Mukerjee}, \citenamefont {Turner},\ and\
  \citenamefont {Moore}}]{PhysRevLett.102.255701}%
  \BibitemOpen
  \bibfield  {author} {\bibinfo {author} {\bibfnamefont {F.}~\bibnamefont
  {Pollmann}}, \bibinfo {author} {\bibfnamefont {S.}~\bibnamefont {Mukerjee}},
  \bibinfo {author} {\bibfnamefont {A.~M.}\ \bibnamefont {Turner}},\ and\
  \bibinfo {author} {\bibfnamefont {J.~E.}\ \bibnamefont {Moore}},\ }\bibfield
  {title} {\bibinfo {title} {Theory of finite-entanglement scaling at
  one-dimensional quantum critical points},\ }\href
  {https://doi.org/10.1103/PhysRevLett.102.255701} {\bibfield  {journal}
  {\bibinfo  {journal} {Phys. Rev. Lett.}\ }\textbf {\bibinfo {volume} {102}},\
  \bibinfo {pages} {255701} (\bibinfo {year} {2009})}\BibitemShut {NoStop}%
\end{thebibliography}
%

\end{document}